\documentstyle{article}
\title{\LARGE Two Poisson structures invariant with respect to discrete transformation in the case of arbitrary semi-simple algebras}
\author{A.~N.~Leznov\thanks{ Universidad Autonoma del Estado de Morelos, CCICAp,Cuernavaca, Mexico}} \date{}

\newcommand{\rig}[2]{\stackrel{#2\rightarrow}{#1}}

\begin{document}
\maketitle

\maketitle

\begin{abstract}
Two Poisson structures invariant with respect to discrete transformation of Maximal root presented in explicit form for arbitrary semi simple algebra. Thus problem of construction of multi-component hierarchies of integrable systems is solved. 
\end{abstract}

\section{Introduction}

In the present paper we would like to systematize results obtained in the previous papers of the author \cite{I},\cite{2} - \cite{7} concerning the discrete transformations, hierarchies of systems of equations and its multi-soliton solutions of $n$ - wave problem. Now we would like to forget about the origin of the problem and consider some general property of semi-simple algebra. The problem may be formulated as the following. With each semi-simple algebra it is possible to connect two Poisson structures invariant with respect to some discrete transformation of the given form. 
Of course this observation was done only after consideration of the examples of the semi simple algebras of the low ranks. And we advise to the reader at first read one of the mentioned above papers in which calculations may be done by fourth action of arithmetic plus differentiation.

\section{Participants of the game and its rules}

\subsection{Notations}

$G$ - arbitrary semi-simple algebra. $R$ - the space of its positive and negative roots.The dimension of $R$ equal $2n=N-r$, $N$ - dimension of the algebra, $r$ its rank. $X_R$ are generators with commutation relations $[X_R,X_R']=D^{R+R'}_{R,R'}X_{R+R'}$ where $D^{R+R'}_{R,R'}$ structure constants (some times we omitted sign $X_R\to R$).  
$f_R$ - the system of functions depending on one parameter $x$ - space coordinate (and possible some other parameters). Some times we use notation $f_{{\pm R}}=f^{\pm}_R$. Elements of Cartan sub algebra will be denoted by little latine letters $c=\sum c_ih_i, [c,X_R]=c_R X_R$. Generators of all roots of the algebra are normalized on unity $(X_R,X_{R'}=\delta_{R,-R'}$.

\subsection{The grading of the maximal root}

In what follows we use the grading of the maximal root \cite{LMP}. This means that
all generators of the algebra may be distributed on the subspaces with $\pm 2,\pm 1$ and $0$grading indexes. The grading index is defined by the proper value of the Cartan generator of the maximal root $H_M=[X^+_M,X^-_M], [H_M,X^{\pm}_M]=\pm 2 X^{\pm}_M$. Thus arbitrary element of semi simple algebra may be represented as
$$
f=f^{(+2}+f^{(+1}+f^0+f^{(-1}+f^{(-2},\quad [H_M,f^{(m}]=m f^{(m}
$$
The subspaces $f^{(\pm 2}=f^{(\pm}_M X^{\pm}_M$ are one dimensional. The subspaces with
$\pm 1$ graded indexes $f^{(\pm 1}=f^{(\pm}_++f^{(\pm}_-$ may be decoupled in such way that 
$[f^{(\pm}_+,f^{(\pm}_-]=cX^{\pm}_M$ and all generators in subspaces with the same additional index $\pm$ are mutually commutative. Elements of of one graded subspaces are represented in a
form $f^{(\pm}=\sum f_{R^{(+1}(S^{(-1})}X_{R^{(+1}(S^{(-1})}$. In what follows zero graded subspace is limited by additional condition $[f^0,X^{\pm}_M]=0$. The zero graded subspace consists from the different elements $[X_{R^{(+1}},X_{S^{(-1}}]$ not coincide with the elements of Cartan sub algebra. The following Furie decomposition take place
$$
\sum (A^{(+1}X_{-R^{(+1}(S^{(-1})})(B^{(-1}X_{R^{(+1}(-S^{(-1})})=(A^{(+1}B^{(-1})
$$
and some modified formula with respect to summation on elements of zero order subspace
$$
\sum (A^0X_{-r_0})(B^0X_{r_0})=(A^0(B^0-\sum h_{\beta}k^{-1}_{\beta,\gamma}
(h_{\gamma}B^0))=
$$
$$
(A^0B^0)-\sum (h_{\beta}B^0)k^{-1}_{\beta,\gamma}(h_{\gamma}A^0)
$$
The last modification is connected with the fact that Cartan elements of the simple roots of the algebra may be among $A^0,B^0$ and they must be excluded from the result of summation on zero order subspace. 

\subsection{Discrete transformation, Frechet derivative and Poisson structure}

We repeat corresponding text from  \cite{DL}.
The discrete invertible substitution (mapping) defined as
\begin{equation}
\tilde u=\phi(u,u',...,u^{r})\equiv \phi(u)\label{1}
\end{equation}
$u$ is $s$ dimensional vector function; $u^r$ its derivatives of corresponding order with 
respect to "space" coordinates.

The property of invertibility means that (\ref{1}) can be resolved and "old" function $u$
may expressed in terms of new one $\tilde u$ and its derivatives.

Frechet derivative $\phi'(u)$ of (\ref{1}) is $s\times s$ matrix operator defined as
\begin{equation}
\phi'(u)=\phi_{u}+\phi_{u'}D+\phi_{u''}D^2+...\label{FR}
\end{equation}
where $D^m$ is operator of m-times differentiation with respect to space coordinates.

Let us consider equation
\begin{equation}
F_n(\phi(u))=\phi'(u)F_n(u)\label{ME}
\end{equation}
where $F_n(u)$ is s-component unknown vector function, each component of which depend on $u$
and its derivatives not more than $n$ order. If equation (\ref{ME}) posses at least one not trivial solution ($F^{trivial}\equiv u'$) then substitution (\ref{1}) is called integrable one.

With each discrete substitution it is possible try to connect Poisson structure defined as
anti symmetric matrix valued operator $J^T(u)=-J(u)$ and its invariance with the respect to discrete transformation (\ref{1}) is fixed by equation it satisfied
\begin{equation}
\phi'(u)J(u)(\phi'(u))^T=J(\phi(u)),\quad \phi'(u)H(u)(\phi'(u))^{-1}=H(\phi(u))\label{H}
\end{equation}
where $(\phi'(u))^T=\phi_{u}^T-D\phi_{u'}^T+D^2(\phi_{u''})^T+...$ and $J(u),H(u)$ are unknown $s\times s$ matrix operators, the matrix elements of which are polynomial of some finite order with respect to operator of differentiation (of its positive and negative degrees). First equation in (\ref{H}) is condition of invariance of Poisson structure with respect to discrete transformation, the second one is equation for raising operator. Two different Poisson structures lead obviously to $H(u)=J(u)_1J^{-1}(u)_2$. 

\section{The statement of the problem}

In \cite{I} it was shown that equations of n-th waves are invariant with respect some number of discrete transformations (coincide with $r$ - the rank of corresponding semi-simple algebra).
Namely with such kind of discrete transformation we will have deal in what follows below.
It will be shown that there exist two different Poisson structures (which can be used for construction of the hierarchies of integrable systems of equations). But the fact of existence
of such Poisson structures is the inner property of semi-simple algebra and the written ad hoc
discrete mapping of the definite form. Author don't know and have no guess how these two objects may be connected in the frame work of group representation theory. 

\subsection{Discrete transformation of the maximal root}

The discrete transformation of the maximal root \cite{LMP},\cite{I} is the mapping realized on the space of $f_R$ functions depending on $r$ arbitrary parameters $[c,X_{R}]=c_R X_{R}$ and one space coordinate $x$ ($\frac{\partial f}{\partial x}\equiv f'$) 
$$
\rig{f_M^+}{T_M}={1\over f_M^-},\quad \rig{f^{(+1}}{T_M}={[X^+_M,f^{(-1}]\over f_M^-}\equiv
{\alpha^{(+1}\over f_M^-},\quad \rig{f_0}{T_M}=(f^0+{1\over 2f_M^-}[\alpha^{(+1},f^{(-1}])_h,
$$
where $(..)_h$ means factor on Cartan sub-algebra. In other words
$$
([\alpha^{(+1},f^{(-1}]])_h\equiv [\alpha^{(+1},f^{(-1}]-\sum h_{\beta}k^{-1}_{\beta,\gamma}
(h_{\gamma}[\alpha^{(+1},f^{(-1}])
$$
where $h_{\gamma}$  Cartan element of simple root of the algebra. 
Below we use the notation $\theta \equiv([X^+_M,f^{(-1}][c,f^{(-1}]))$
$$
[c,\rig{f^{(-1}}{T_M}]=(f^{(-1})'+{(f^-_M)'-{1\over 2}\theta \over c_M f^-_M}[c,f^{(-1}]+
{1\over 6f^-_M}[c[[\alpha^{(+1},f^{(-1}]f^{(-1}]]-
$$
$$
[f^{(-1},[c,f^0]]+f_M^-[[c,f^{(+1}]X^-_M]
$$
$$
c_M\rig{f_M^-}{T_M}=c_M(f_M^-)^2f^+_M+{(f_M^-)''\over c_M}+{1\over c_M}([X^+_M,f^{(-1}]
[c,(f^{(-1})'])-{((f_M^-)'-{1\over 2}\theta)^2\over c_M f_M^-}
$$
\begin{equation}
+{1\over 24 f_M^-}([\alpha^{(+1},f^{(-1}][\alpha^{(+1},f^{(-1}])
+f_M^-(f^{(-1}[c,f^{(+1}])-{1\over 2}([\alpha^{(+1},f^{(-1}][c,f^0])\label{BIG}
\end{equation}

We would like to show that transformation (\ref{BIG}) is canonical one. That means that there exist some functional $G$ (generating function) depending on the pairs of transformed and initial variables $\rig{f_R}{T_M},f_{R'}$ in which encoded all information about (\ref{BIG}). We will use the following identification for general impulses $ P^M=c_Mf^+_M,P^1=[c,f^{(+1}],P^0=[c,f^0_+]$ and coordinates $ X_M=f^-_M,X_1=f^{(-1},X_0=f^0_-$ where $f^0_{\pm}$ means part of $f^0$ connected with positive and negative spaces of zero order subspace. We choose the following dependence of mentioned above functional $G=G(X_{M,1},\rig{X_{M,1}}{T_M}; X_0,\rig{P_0}{T_M})$ then in the case of canonical transformation
\begin{equation}
P^{M,1}=-\frac{\delta G}{\delta X_{M,1}},\quad \rig{P^{M,1}}{T_M}=\frac{\delta G}{\delta \rig{X_{M,1}}{T_M}},\quad P^0=-\frac{\delta G}{\delta X_0},\quad \rig{X_0}{T_M}=-\frac{\delta F}{\delta \rig{P^0}{T_M}}\label{CT}
\end{equation}
where functional derivatives coincide with the Frechet one (\ref{FR}).
In the case of canonical transformation of the classical dynamics (and thermodynamic)
functional $G$ depend only on canonical coordinates and generalized impulses but not on its derivatives. It is not surprising that discrete transformation is connected with generalized canonical ones. This fact was observed  some times ago and proof of this fact with respect to self-dual system ( and many other integrable systems) reader can find in \cite{LR}.

We present below explicit expression for density of functional $G$. Reader can check by using only one operation of differentiation that (\ref{CT}) and (\ref{BIG}) are identical
$$
G=c_M{(\rig{f^-_M}{T_M})\over f^-_M})-{1\over f^-_M}(\alpha^{(+1}[c,\rig{f^{(-1}}{T_M}])-{1\over 2c_M}({f'_M\over f_M})^2-{1\over c_Mf^-_M}(\alpha^{(+1}[c,(f^{(-1})'])-
$$
$$
-{1\over 8c_M(f^-_M)^2}(\alpha^{(+1}[c,f^{(-1}])^2+{1\over 48(f^-_M)^2}([\alpha^{(+1},f^{(-1}][\alpha^{(+1},f^{(-1}])-
$$
\begin{equation}
(\rig{P^0}{T_M}X_0)+{1\over 2f^-_M}([c[\alpha^{(+1},f^{(-1}]]_+X_0)-\label{FUNK}
\end{equation}
$$
-{1\over 2f^-_M}([\alpha^{(+1},f^{(-1}]_-\rig{P^0}{T_M})+{1\over 4f^-_M}([c[\alpha^{(+1},f^{(-1}]]_+[\alpha^{(+1},f^{(-1}]_-)
$$
As reader can see from (\ref{FUNK}) generating function in the case under consideration depend 
on the derivatives of the first order of $f^{(-1}_R$ functions and second order on $f^-_M$ one.
It lead to some modification in usual theory of canonical transformation which will be noticed below. 

\section{Conservation of the Poisson brackets}

The main property of canonical transformation consists in conservation of Poisson brackets between dynamical variables involved. This fact proofs automatically in the usual theory (without derivatives of dynamical variables). We will try to repeat corresponding calculation
and show that this property with respect to discrete transformation of the previous section is conserved.

For this goal let us consider  Jacobi matrix
\begin{equation}
J={D(\rig{P}{T_M}, \rig{X_0}{T_M}; \rig{P_0}{T_M}, \rig{X}{T_M})\over D( p,x_0;p_0,x)}=
{D( \rig{P}{T_M}, \rig{X_0}{T_M}; \rig{P_0}{T_M}, \rig{X}{T_M})\over D( p,p_0;x_0,x)}\pmatrix{1 & 0 & 0 & 0 \cr
0 & 0 & 1 & 0 \cr
0 & 1 & 0 & 0 \cr
0 & 0 & 0 & 1 \cr}\label{JD}
\end{equation}
In (\ref{JD}) in $P$ we unite generalized impulses $P$=$P_M,P^1$ and the same do with corresponding coordinates $X$=$X_M,X^1$. We can not do the same with coordinates and impulses with zero indexes by the same reason why unite canonical transformation in usual theory my be written not in all pairs of canonical variables. 
For further transformation of (\ref{JD}) we unite "4" independent variables of $G$ functional into two pairs
$\bar y=(\rig{P_0}{T_M}, \rig{X}{T_M})$, $y=(x,x_0)$. In this notations (\ref{CT}) can be rewritten as follows
$$    
\pmatrix{ \rig{P}{T_M}\cr
         \rig{X_0}{T_M} \cr}=\pmatrix{ 0 & 1 \cr
                                     -1 & 0 \cr}G_{\bar y}\equiv \sigma_2 G_{\bar y},\quad \pmatrix{p \cr
           p_0 \cr}=-G_y
$$
After differentiation of the last equality with respect to pairs of arguments $p=(p,p_0),y=(x,x_0)$
and taking into account independent arguments of generating function $G$ we obtain
$$
\bar y_{p}= -(G_{y,\bar y})^{-1},\bar y_x=-(G_{y,\bar y})^{-1}G_{y,y},\quad
$$
$$
\pmatrix{ \rig{P}{T_M}\cr
\rig{X_0}{T_M} \cr}_p=-\sigma_2 G_{\bar y,\bar y}(G_{y,\bar y})^{-1},
$$
$$
\pmatrix{ \rig{P}{T_M}\cr
         \rig{X_0}{T_M} \cr}_y=-\sigma_2 G_{\bar y,y}\sigma_1+\sigma_2 G_{\bar y,\bar y}(G_{y,\bar y})^{-1}G_{y, y}
$$
$$
(\sigma_1=\pmatrix{0 & 1 \cr
                  1 & 0 \cr},\quad \sigma_2=\pmatrix{0 & 1 \cr
                                                     -1 & 0 \cr},\quad 
\sigma_3=\pmatrix{1 & 0 \cr
                  0 & -1 \cr}).
$$
In the general case, when generating functional has arbitrary dependence on derivatives of $p,x$ arguments the last expression may be had no mean, but in the case under consideration generating functional is linear in $\bar y$ arguments. In this case matrix $G_{y,\bar y}$ depend only on generalized coordinates $X$ and impulses $P$ but not from their derivatives and all operations done above are absolutely correct.
After this manipulations expression for Jacobi matrix valued operator may rewritten in two dimensional form ( we take into account condition of linearity of $G,G_{\bar y,\bar y}=0$
$$
J=\pmatrix{ 0 & \sigma_3 G_{\bar y,y}\sigma_1 \cr
 -(G_{y,\bar y})^{-1} & -(G_{y,\bar y})^{-1}G_{y,y}\sigma_1 \cr}\pmatrix{1 & 0 & 0 & 0 \cr
0 & 0 & 1 & 0 \cr
0 & 1 & 0 & 0 \cr
0 & 0 & 0 & 1 \cr}(=\delta (2\to 3))
$$
Now by direct calculations let us check that $J \pmatrix{ 0 & \sigma_3 \cr
                                                         \sigma_3 & 0 \cr}J^T=
                                                     \pmatrix{ 0 
& \sigma_2 \cr                                                    
                                                         \sigma_2 & 0 \cr}$.
Indeed 
$$
\sigma_1\delta (2\to 3)\pmatrix{ 0 & \sigma_3 \cr                                                    
                  -\sigma_3 & 0 \cr}\delta (2\to 3)^T\sigma_1^T=\pmatrix{ 0 & I \cr                                                    
                                                -I & 0 \cr}
$$ 
and
$$
J\pmatrix{ 0 & I \cr                                                    
          -I & 0 \cr}J^T=\pmatrix{ 0 & \sigma_3 G_{\bar y,y} \cr
 -(G_{y,\bar y})^{-1} & -(G_{y,\bar y})^{-1}G_{y,y} \cr}\pmatrix{ G_{y,\bar y}\sigma_3 & G^T_{y,y}(G_{\bar y,y})^{-1} \cr
 0 & -(G_{\bar y,y})^{-1} \cr}=
$$
$$
\pmatrix{ 0 & \sigma_3 \cr
        -\sigma_3 & G_{y,\bar y})^{-1}(G^T_{y,y}-G_{y,y})(G_{\bar y,y})^{-1} \cr}
$$
In the case when $G$ don't depend on space derivatives the term in the right dawn corner of the last matrix obviously equal to zero. In the case of functional (\ref{FUNK}) the checking of this fact reader find in Appendix.

In the previous papers it was used the order of functions $f^+_M,f^{(+1},f^0_+,f^0_,f^{(-1},f^+_M$
To come to such order from considered above it is necessary to do some number of point like tangent transformation: change by the places $X_0\to P_0$ and do mirror reflection in $X_0,X$ terms. After such transformations the zero degree Poisson structure looks as
$$
\pmatrix{ 0 & \sigma \cr
-\sigma & 0 \cr}
$$
where $\sigma$ is matrix $(1+N^1+N^0\times 1+N^1+N^0)$ with different from zero unity on its main anti diagonal.

The proof of conservation of the Poisson brackets may be obtained more directly from the fact of
existence of Lagrangian function \cite{LEZ4} ( and action for n-wave system)
$$
L(f)=Sp {1\over 2}([d,f_t]+[c,f_x])+{1\over 3}(f[[d,f][c,f]]), S=\int d t d x L(t,x)
$$
change on the derivatives under transformation of maximal root above.

\subsection{Frechet derivative}

The Frechet derivative of (\ref{BIG}) in connection of its definition (\ref{FR}) may be presented in block form as $5\times 5$ matrix valued operator
\begin{equation}
\pmatrix{ 0 & 0 & 0 & 0 & \phi'_{M,-M} \cr
0 & 0 & 0 & \phi'_{R^{(+1},S^{(-1}} & \phi'_{R^{(+1},-M} \cr
0 & 0 & \phi'_{r_0,r'_0} & \phi'_{r_0,S^{(-1}} & \phi'_{r_0,-M} \cr
0 & \phi'_{S^{(-1},R^{+1}} & \phi'_{S^{(-1},r_0} & \phi'_{ S^{(-1}, (S^{(-1})'} & \phi'_{S^{(-1},-M} \cr
\phi'_{M,-M} & \phi'_{-M,R^{+1}} & \phi'_{-M,r_0} & \phi'_{-M, S^{(-1}} & \phi'_{-M,-M} \cr}\label{FRCT}
\end{equation}
where $I$ is the unity matrix with the dimension of zero grading subspace: the notation $M_{R}$
means that from the algebra valued function $M$ necessary extract coefficient under corresponding root $R$ of the algebra belonging to grading spaces with indexes $\pm 2,\pm 1$ and $0$.
$$
\phi'_{M,-M}=-{1\over (f^-_M)^2},\quad \phi'_{{R^{(+1}}, S^{(-1}}={([X^+_M,X_{S^{(-1}}]X_{-R^{(+1}})\over f^-_M},
$$
$$
\phi'_{{R^{(+1}},-M}=-{([X^+_M,f^{(-1}]X_{-R^{(+1}})\over (f^-_M)^2},\quad \phi'_{r_0,r'_0}=I,\quad \phi'_{r_0,S^{(-1}}={([[X^+_M,X_{S^{(-1}}]f^{(-1}]X_{-r^0})\over f^-_M}
$$
$$
 \phi'_{r_0,-M}=-{1\over 2}{([[X^+_M,f^{(-1}]f^{(-1}]X_{r^0})\over (f^-_M)^2},\quad \phi'_{S^{(-1},R^{+1}}={f^-_M\over c_{S^{(-1}}}([[c,X_{R^{+1}}]X^-_M]X_{-S^{(-1}}),X
$$
$$
\phi'_{S^{(-1},r_0}=-{1\over c_{S^{(-1}}}([f^{(-1}[c,X_{r^0}]]X_{-S^{(-1}}),\quad \phi'_{-M,M}=(f^-_M)^2,\quad \phi'_{-M,R^{(+1}}=-{f^-_M\over c_M}(f^{(-1}[c,X_{R^{(+1}}])
$$
$$
\phi'_{-M,r_0}={1\over 2c_M}([[X^+_M,f^{(-1}]f^{(-1}][c,X_{r^0}]),\quad \phi'_{-M, S^{(-1}}={c_{ S^{(-1}}\over c^2_M}([X^+_M,f^{(-1}]X_{ S^{(-1}})D+
$$
$$
{1\over 6c_M f^-_M}([X^+_M,f^{(-1}]X_{ S^{(-1}}][[X^+_M,f^{(-1}]f^{(-1}])+{f^-_M\over c_M}([c,f^{(+1}]X_{ S^{(-1}})-{1\over c_M}([[X^+_M,X_{ S^{(-1}}]f^{(-1}][c,f^0])-
$$
$$
{1\over (c_M)^2}([[X^+_M,X_{ S^{(-1}}][c,(f^{(-1})'])+(2c_{S^{(-1}}+c_M){(f^-_M)'-{1\over 2}([X^+_M,f^{(-1}]([c,f^{(-1}])\over c^2_M f^-_M}([X^+_M,f^{(-1}]X_{ S^{(-1}})
$$
$$
\phi'_{S^{(-1},-M}={f_{ S^{(-1}}\over f^-_M c_M}D-{f_{ S^{(-1}}\over (f^-_M)^2 c_M}(f^-_M)'-
{1\over 6(f^-_M)^2}(X_{-S^{(-1}}[[[X^+_M,f^{(-1}]f^{(-1}]f^{(-1}])+
$$
$$
{f_{ S^{(-1}}\over 2c_M(f^-_M)^2}{1\over c_M})([X^+_M,f^{(-1}][c,f^{(-1}])+
{1\over c_{S^{(-1}}}([X^-_M,X_{ -S^{(-1}}][c,f^{(+1}])
$$
$$
\phi'_{S^{(-1},(S^{(-1})'}=\delta_{S^{(-1},(S^{(-1})'}({1\over c_{S^{(-1}}}D+{(f^-_M)'-{1\over 2}\Theta \over c_M f^-_M})-{c_{(S^{(-1})'}f_{S^{(-1}}\over c_M f^-_M}(X_{(S^{(-1})'}[X^+_M,f^{(-1}])-
$$
$$
{1\over c_{S^{(-1}}}(X_{-S^{(-1}}[X_{(S^{(-1})'}[c,f^0]])+
{1\over 2f^-_M}(X_{-S^{(-1}}[[\alpha^{(+1},f^{(-1}]X_{(S^{(-1})'}]
$$
$$
\phi'_{-M,-M}={1\over (c_M)^2}D^2-2{((f_M^-)'-{1\over 2}\Theta)\over (c_M)^2 f_M^-}D+
2(f_M^-f_M^++{((f_M^-)'-{1\over 2}\Theta)^2\over (c_M)^2 (f_M^-)^2}-
$$
$$
{1\over 24 (f_M^-)^2}([\alpha^{(+1},f^{(-1}][\alpha^{(+1},f^{(-1}])
-{1\over c_M}(f^{(-1}[c,f^{(+1}])
$$

\section{The main assertion}

There are two Poisson structures $J_0$ and $J_1$, which are invariant with respect to discrete transformation of the maximal root of the previous subsection (equations (\ref{H})
are satisfied) with the explicit form of its matrix elements
\begin{equation}
<R| J_0 |R'>=\delta_{R,-R'}{1\over c_R}, c_{-R}=-c_R \label{P0}
\end{equation}
The fact of invariance of $J_0$ Poisson structure with respect to discrete transformation was proofed in the previous section.
\begin{equation}
<R| J_1 |R'>=-f_R(h_R,h_{R'}) D^{-1} f_{R'}+{1\over c_{R}}[(X_{-R},X_{-R'}][c,f]){1\over c_{R'}}+{1\over c_R^2}\delta_{R,-R'}D\label{JAp}
\end{equation}
where $h_R=\sum n^R_{\alpha} h_{\alpha}$, $n^R_{\alpha}$ multiplicity simple root $\pm \alpha$
in the root $R$ and  $ h_{\alpha}$ Cartan element of this simple root.
The proof of the last (not so trivial) proposition will be presented below.

\subsection{Simplest nontrivial solution of (\ref{ME})}

The column function $F_{R}=g_R f_R$ is the solution of the equation (\ref{ME}). We have
$$
(\phi'(f)F)|R>=\sum <R | \phi'(f) |R'>F_{R'}=\sum <R | \phi'(f) |R'> g_{R'} f_{R'}=
$$
$$
\sum \frac{\partial \rig{f_R}{T_M}}{\partial f_{R'}}[g,f_{R'}]=[g,\rig{f_R}{T_M}]=g_R\rig{f_R}{T_M}
$$
Indeed $g_R f_R$ may be considered as result of differentiation $[g,f_{R}]$ and the discrete transformation (\ref{BIG}) conserve the grading structure with respect to simple roots of the algebra.

\subsection{Equation which it is necessary to check}

Let us rewrite the equation for Poisson structure $J$ (\ref{H}) taking into account the fact of existence of $J_0$ (\ref{P0}). We have $(J^T=-J)$
\begin{equation}
\phi'(f)J=\rig{J}{T_M}(\phi'(f)^{-1})^T=-(\phi'(f)^{-1}\rig{J}{T_M})^T=-(J_0\phi'(f)^T J_0^{-1}\rig{J}{T_M})^T\label{JH} 
\end{equation}
Let us present first term in (\ref{JAp}) in equivalent form
$$
f_R(h_R,h_{R'}) D^{-1} f_{R'}=\sum_{i=1}^r F_i D^{-1} F_i^T
$$
where $F_i=g^i_R f_R$ are $r$ independent nontrivial solutions of the first subsection.
Then the form above is equivalent to (\ref{JAp}) after identification $((g_R,g_{R'}) \equiv \sum_ig^i_R g^i_{R'}=(h_R,h_{R'})$. And this fact it will necessary to prove. Now let us consider result of multiplication Frechet derivative on (column) nontrivial solution $F_i$.
Using the rule of multiplication of quadratical in derivatives operator on scalar function $F$ 
\begin{equation}
(A+B D+C D^2)=A F+B F'+C F''+(B F+2C F')D+C F D^2\label{R}
\end{equation}
and explicit form of Frechet derivative we conclude that first $1+n^1+n^0$ lines of it does not contain operator of differentiation $D$, next $n^1$ ones are linear in $D$ and the last one
is quadratical in it (we remind that 1-dimension of $\pm 2$ graded subspaces, $n^1,n^0$ dimensions of $\pm 1$,$0$ ones). We present below $n^1+1\times n^1+1$ matrix in the left dawn corner of Frechet derivative which contain operator of differentiation in explicit form
\begin{equation}   
\pmatrix{ {1\over c_{S^{-1}}}\delta_{(S^{-1})',S^{-1}}D & {f_{S^{-1}}\over c_M f^-_M} D \cr
{1\over c_M^2}([c,X_{(S^{-1})'}][X^+_M,f^{(-1}])D & {1\over c_M^2}D^2-2{((f^-_M)'-{1\over 2}([c,f^{(-1}][X^+_M,f^{(-1}]))\over c_M^2f^-_M}D \cr}\label{RDC}
\end{equation}
In connection with (\ref{ME}) $\rig{F_i}{T_M}=\phi(f) F_i$
(in the last formulae all operators of differentiation acts directly on $F_i$). And thus we have
\begin{equation}
\phi(f) F_i=\rig{F_i}{T_M}+\pmatrix{ 0_{1+n^1+n^0} \cr   
-({g^i_M\over c_M}-{g^i_{S^{-1}}\over c_{S^{-1}}})f_{S^{-1}}D \cr
-{g^i_M f^-_M\over c_M^2}D^2+{1\over c_M^2}([c,f^{(-1}][X^+_M,[g^i ,f^{(-1}]]) D \cr}\equiv  \label{BITL}  
\end{equation}
$$
\rig{F_i}{T_M}+\Gamma_i
$$
The assumed form of second Poisson structure contain terms with negative, zero and positive 
degrees of $D$ ($J^{-1}_1,J^0_1,J^{1_1}$). The (\ref{BITL}) together with (\ref{JH}) and comments after leads to conclusion
$$
\phi'(f)J^{-1}_1=-\sum_{i=1}^r (\rig{F_i}{T_M} D^{-1} F_i^T+\Gamma_i D^{-1} F_i^T)
$$
But $\Gamma_i$ is proportional to $D$ and thus last term in the sum above contain only $0,1$ degrees of $D$.
Thus explicit expression for  $-\sum_{i=1}^r(\Gamma_i D^{-1} F_i^T)$ looks as
$$
\pmatrix{ 0_{1+n^1+n^0} \cr   
({(M,P)\over c_M}-{(S^{-1},P)\over c_{S^{-1}}})f_{S^{-1}} \cr
{(M,P) f^-_M\over c_M^2}D-{1\over c_M^2}([c,f^{(-1}][X^+_M,[\sum (g^i g^i_P) ,f^{(-1}]]) \cr} (F_P)^T
$$ 
In the last column different from zero only components of its $n_1+1$ last rows.
In what follows we use notation $\Theta=([c,f^{(-1}][X^+_M,f^{(-1}]])$, 
$\Theta_P=([c,f^{(-1}][X^+_M,[\sum (g^i g^i_P) ,f^{(-1}]])$.
Now let us inverse (\ref{BITL}) presenting it in a form
$$
F_i^T=(\rig{F_i}{T_M})^T (\phi^{-1}(f))^T+\Gamma_i^T(\phi^{-1}(f))^T
$$

For calculation $(\phi'(f)^{-1})^T=J_0^{-1}\phi'(f)J_0$ it is suitable to present $J_0$ as product of two matrices one diagonal one with matrix elements $ <R| d |R'>=\delta_{R,R'}{1\over c_R}$ and the second with different from zero unites on its main anti diagonal $<R| ad |R'>=\delta_{R,-R'}$. Thus we obtain $(\phi'(f)^{-1})^T=J_0^{-1}\phi'(f)J_0$=
\begin{equation}
d^{-1}\pmatrix{ \phi'_{-M,-M} & \phi'_{-M,-(R^{(+1})'} & \phi'_{-M,-r'_0} & 
\phi'_{-M,-(S^{(-1})'} & \phi'_{-M,M} \cr
\phi'_{-R^{(+1},-M} & \phi'_{-R^{(+1},-(R^{(+1})'} & \phi'_{-R^{(+1},-r'_0} & \phi'_{-R^{(+1},-(S^{(-1})'} & 0 \cr
\phi'_{-r_0,-M} & \phi'_{-r_0,(-R^{(+1})'} & I & 0 & 0 \cr
\phi'_{-S^{(-1},-M} & \phi'_{-S^{(-1},-(R^{(+1})'} & 0 & 0 & 0 \cr
\phi'_{M,-M} & 0 & 0 & 0 & 0 \cr}d \label{FRCII}
\end{equation}
Now calculation of $\sum \rig{F_i}{T_M} D^{-1}\tilde \Gamma_i^T=\sum \rig{F_i}{T_M} D^{-1}\Gamma_i^T(\phi^{-1}(f))^T$ became obvious with the result
\begin{equation}
\rig{F_i}{T_M}(\pmatrix{-{(PM) \over c_M^2}D{1\over (f^-_M)}+{1\over (f^-_M)^2c_M^2}\Theta_P &  
({PM\over c_M}-{(P R^{+1})\over c_{R^{+1}}})\rig{f_{R^{+1}}}{T_M} &  0_{1+n^1+n^0} \cr}) 
\label{BITL2}
\end{equation}
In the last line only first $(1+n_1)$ terms are different from zero.

Substituting all results obtained above into main equation (\ref{JH}) we come to equality which have to be checked
$$
-\sum \Gamma_i D^{-1} F_i^T+\phi'(f)J^0_1+\phi'(f)J^1_1=
$$
\begin{equation}
\sum \rig{F_i}{T_M} D^{-1} \tilde \Gamma_i +\rig{J^0_1}{T_M}(\phi'(f)^{-1})^T+\rig{J^1_1}{T_M}(\phi'(f)^{-1})^T \label{JHF} 
\end{equation}
We pay attention the reader that in (\ref{JHF}) the terms containing $D^{-1}$ mutually cancel and all terms in equality above contain only positive degree of operator $D$.
The equation (\ref{JHF}) is equation on matrix elements of $(2+2N^1+N^0)\times (2+2N^1+N^0)$ matrix. We present this matrix as a sum of 4 matrix quadratic $(1+N^1+N^0\times 1+N^1+N^0)$ one in its upper right corner, two rectangular $(1+N^1\times 1+N^1+N^0)$ and $(1+N^1+N^0\times 1+N^1)$ in its upper left and right dawn corners correspondingly and quadratical $(1+N^1\times 1+N^1)$ in its left dawn corner. Below we prove that all matrix elements of left and right sides of (\ref{JHF}) are equal to each other. Of course this calculations are not simple one and author absolutely sure that they are not necessary but at the present moment have no idea, on what it is possible to change them.

\section{Proofs of validity of the main equation (\ref{JHF})}

Some general comments. In all calculations we will have deal with matrix elements of product of two  matrices. The sum on mediate states always has a form $\sum_{P} c^{\nu}_P(K,X_{-P})(L,X_P),\nu=-1,0$. When $\nu=0$ the summation may be performed in explicit with help the comments in the second section. In the case $ \nu=-1$ it is necessary find the same terms in both side of equality to be checked. The strategy of calculations below will be the same. At first we consider coefficient before positive degree of $D$. In nonlocal sums we find terms of the same structure in both sides. All non locality cancels. And summation it is possible perform in explicit form. In some cases under consideration below one of the factors in the expression under summation fixed uniquely value $\nu_P$ and thus it is possible to take this factor out of the sign of the sum with further summation in explicit form.

\subsection{Checking of the elements of $(1+N^1+N^0\times 1+N^1+N^0)$ quadratical matrix of right upper corner of (\ref{JHF})}

For these elements $\Gamma_i=\tilde \Gamma_j=0$ and the main equation (\ref{JHF}) looks as
(we introduce additional notation $J^p_1\equiv J^0_1+J^1_1$)
$$
\phi'(f)J^p_1=\rig{J^p_1}{T_M}((\phi'(f)^{-1})^T
$$
0r in more detail form
$$
[\sum_{P}{\phi'_{R,P}X_{-P}\over c_P},X_{-R'}]{1\over c_{R'}}={1\over c_{R}}[X_{-R},\sum_{P}X_{-P}((\phi')^{-1})^T_{R,P}]{1\over c_{R'}}
$$
where $R=M,R^{(+1},r_0, R'=-M,S^{-1},r_0$.

Below we present calculations for more complicate case considering the last equality between the states $r_0,r_0'$. 
$$
(J^p_1)_{r_0,r_0'}+\sum_{S^{-1}}\phi'_{r_0,S^{-1}}(J^0_1)_{S^{-1},r_0'}=(\rig{J^p_1}{T_M})_{r_0,r_0'}+\sum_{R^{(+1}}(\rig{J^p_1}{T_M})_{r_0,R^{(+1}}((\phi'(f)^{-1})^T)_{R^{(+1},r_0'}
$$ 
In writing the last equality we conserve all not vanishing terms in connection with definition of Frechet derivative, assumed form of $J^p_1$ and the fact that $\phi'(f)_{r_0,r_0'}$ is unity matrix. Now terms with derivatives in $(J^1_1)_{r_0,r_0'}=\delta_{r_0,-r_0'}D$ does not depend from shifts and thus cancels in both sides of equality. Further
$$    
(\rig{J^p_1}{T_M})_{r_0,r_0'}-(J^p_1)_{r_0,r_0'}={1\over c_{r_0}c_{r'_0}}
([X_{-r_0},X_{-r'_0}][c,(\rig{f^0}{T_M}-f^0)])=
$$
$$
{1\over 2f^-_M c_{r_0}c_{r'_0}}([X_{-r_0},X_{-r'_0}][c,[[X^+_M,f^{(-1}]f^{(-1}])
$$

Now we consider sum in left hand side. Commutator $[X_{-S^{-1}},X_{-r'_0}]$  belongs to $+1$ graded subspace and thus 
$([X_{-S^{-1}},X_{-r'_0}][c,f])=(c_{S^{-1}}+c_{r'_0})([X_{-S^{-1}},X_{-r'_0}]f^{(-1})$. Using this fact and substituting explicit expression for elements of Frechet derivative we rewrite this sum in a form
$$
{1\over f^-_M}\sum_{S^{-1}}({1\over c_{S^{-1}}}+{1\over c_{r'_0}})([X_{-S^{-1}},X_{-r'_0}]
f^{(-1})([[X^+_M,f^{(-1}]X_{S^{-1}}]X_{-r_0})
$$
In corresponding calculations in right side it is necessary to take into account that 
commutator $[X_{-r_0},X_{-R^{+1}}]$ belongs to $-1$ graded subspace and thus 
$$
([X_{-r_0},X_{-R^{+1}}][c,\rig{f}{T_M}])=(-c_{-R^{+1}}+c_{r_0})([X_{-r_0},X_{R^{+1}}]
\rig{f^{(+1}}{T_M})
$$. 
Using this fact, the value for $\rig{f^{(+1}}{T_M}={1\over f^-_M}
[X^+_M,f^{(-1}]$ from (\ref{BIG}) and explicit expression for matrix element of Frechet derivative we present the sum under consideration in a form (in all places we exchange 
$-R^{+1}\to S^{-1}$ to have the same notations as in the sum in the left side calculation above) 
$$
{1\over f^-_M}\sum_{S^{-1}}({1\over c_{S^{-1}}}-{1\over c_{r_0}})([X_{-S^{-1}},X_{-r'_0}]
f^{(-1})([[X^+_M,f^{(-1}]X_{S^{-1}}]X_{-r_0})
$$
The terms containing $c_{S^{-1}}$ in denominator are the same in both sides of
equality and mutually cancels. In remaining terms it is possible perform summation in explicit form  
$$
-{1\over f^-_M}({1\over c_{r_0'}}+{1\over c_{r_0}})
([[X^+_M,f^{(-1}][X_{-r'_0},f^{(-1})]]X_{-r_0})
$$
The last term (after simple manipulations) cancels with  the same obtained on the first step of computation above. Absolutely in the same manner it is possible to convince that equation 
(\ref{JHF}) is satisfied for all elements of the matrix of the left upper corner.
 
\subsection{Checking of the elements of $(1+N^1\times 1+N^1+N^0)$ rectangular matrix of left upper corner of (\ref{JHF})}

For this elements $\Gamma_i=0$, but terms with $\tilde \Gamma_{M},\tilde \Gamma_{R^{(+1}}$ will give input into left hand side of (\ref{JHF}). 

Below we present result of computation the matrix elements of two second terms of left-hand side of (\ref{JHF}) and three terms of right hand side of this equation.

\subsubsection{Matrix element between states $<M|,|M>$}

$$
\sum_P  \phi'_{M,P}J_{P,M}=\phi'_{M,-M}J_{-M,M}=-{1\over c^2_M (f^-_M)^2} D
$$
$$
\rig{f^+_M}{T_M}\sum (g^i_M \tilde \Gamma^i_{M})=
$$
$$
-{(M,M)\over c^2_M}{1\over f^-_M}D{1\over f^-_M}+{1\over c^2_M (f^-_M)^3}([c,f^{(-1}][X^+_M,[\sum (g^i_M g^i),f^{(-1}]])
$$
and at last
$$
\sum_P (\rig{J}{T_M})_{M,P} ((\phi')^{-1})^T_{P,M}=(\rig{J}{T_M})_{M,-M} ((\phi')^{-1})^T_{-M,M}+
$$
$$
(\rig{J}{T_M})_{M,S^{-1}} ((\phi')^{-1})^T_{S^{-1},M}=D{1\over c^2_M (f^-_M)^2}+{1\over c^2_M (f^-_M)^3}([c,f^{(-1}][X^+_M,f^{(-1}])
$$
The terms in the first line and sum in the others are equal  under the condition 
$(M,M)\equiv \sum (g^i_M g^i_M)=2,(M,S^{-1})\equiv \sum (g^i_M g^i_{S^{-1}})=-1$. This is exactly proposition of the main assertion.

\subsubsection{Matrix element between states $<M|,|R^{(+1}>$}

$$
\phi'_{M,-M}J_{-M,R^{(+1}}=-{1\over c_M c_{R^{(+1}}(f^-_M)^2}([[X^+_M,X_{-R^{(+1}}][c,f^{-1}])= 
$$
$$
{(c_M -c_{R^{(+1}})\over c_M c_{R^{(+1}}(f^-_M)^2}([X^+_M,f^{(-1}]X_{R^{(+1}})
$$
$$
-\sum (g^i_M \tilde \Gamma^i_{R^{(+1}})=
$$
$$
\rig{f^+_M}{T_M}(-{(MM)\over c_M}+{(MR^{(+1})\over c_{R^{(+1}}})\rig{f_{R^{(+1}}}{T_M}=(-{2\over c_M}+{1\over c_{R^{(+1}}}){([X^+_M,f^{(-1}]X_{-R^{(+1}})\over (f^-_M)^2}
$$
And at last
$$
\sum_P (\rig{J}{T_M})_{M,P} ((\phi')^{-1})^T_{P,R^{(+1}}=(\rig{J}{T_M})_{M,S^{(-1}} ((\phi')^{-1})^T_{S^{(-1},R^{(+1}}={([X^+_M,f^{(-1}]X_{-R^{(+1}})\over c_M(f^-_M)^2}
$$
The sum of two last rows exactly equal to the first one and thus the main equality (\ref{JHF}) for matrix element under consideration is satisfied.

\subsubsection{Matrix element between states $<R^{(+1}|,|M>$}

$$
\sum_{S^{(-1}}\phi'_{R^{(+1},S^{(-1}}J_{S^{(-1},M}+\phi'_{R^{(+1},-M}J_{-M,M}=\rig{f_{R^{(+1}}}{T_M}(-{(R^{(+1},M)\over c^2_M}D{1\over f^-_M}+
$$
$$
{1\over (f^-_M)^2c^2_M}([c,f^{(-1}][X^+_M,[\sum (g^i_{R^{(+1}} g^i),f^{(-1}]]))-
\rig{J_1}{T_M}_{R^{(+1},-M}\phi'(f)_{M,-M}+
$$
$$
\sum_{S^{(-1}}\rig{J_1}{T_M}_{R^{(+1},S^{(-1}}{c_{S^{(-1}}\over c_M}\phi'(f)_{-S^{(-1},-M}+
\sum_{r_0}\rig{J_1}{T_M}_{R^{(+1},r_0}){c_{r_0}\over c_M}\phi'_{-r_0,-M}+
$$
$$
\sum_{P^{(+1}}\rig{J_1}{T_M}_{R^{(+1},P^{(+1}}){c_{P^{(+1}}\over c_M}\phi'(f)_{-P^{(+1},-M}
$$
In writing of the last equality we exchange matrix elements of $((\phi'(f)^{-1})^T$ on matrix elements of $\phi'(f)$ in connection with  (\ref{FRCII}). 

Now we substitute explicit expressions for transformed values (\ref{FRCII}) and Frechet derivatives and after summation rewrite the last expression in a equivalent form
$$
-{\rig{f_{R^{(+1}}}{T_M}\over f^-_M c^2_M}D-{c_{R^{(+1}}\over c_{M-R^{(+1}}c_M}{f_{R^{(+1}}\over f^-_M}=\rig{f_{R^{(+1}}}{T_M}(-{1\over c^2_M f^-_M}(D-{(f^-_M)'\over (f^-_M)^2})+
$$
$$
{1\over (f^-_M)^2c^2_M}\theta_{R^{(+1}}-{1\over c_{R^{(+1}c_M}}[{1\over (f^-_M)^2}([X_{-R^{(+1}},X^+_M][(f^{(-1})'+
$$
$$
{1\over c_M}{(f^-_M)'-{1\over 2}\theta \over f^-_M}[c,f^{(-1}]+
{1\over 6f^-_M}[c,[[\alpha^{(+1},f^{(-1}]f^{(-1}]])-[f^{(-1}[c,f^0]])]+f^-_M([c,f^{(+1}]X^-_M)-
$$
$$
{1\over c_M}{(f^-_M)'-{1\over 2}\theta \over f^-_M}[c,f^{(-1}]+{1\over  c_{R^{(+1}}c_M (f^-_M)^2}(X_{-R^{(+1}}[\alpha^{(+1}[c,\rig{f^0}{T_M})]])-
$$
$$
{1\over  2c_{R^{(+1}}c_M(f^-_M)^2}(X_{-R^{(+1}}[([\alpha^{(+1},f^{(-1}])_h][c,\rig{f^{+1}}{T_M}]])-
$$
$$
{1\over c_{R^{(+1}}f^-_M}([X_{-R^{(+1}},X_{-M+R^{(+1}}]X^+_M)(\phi'(f))_{-M+S^{(-1},-M}]
$$
The both sides of the last expression contain terms first and zero degree in $D$. The terms linear in derivative are the following one ($(R^{(+1}M)=1$)
$$
-{\rig{f_{R^{(+1}}}{T_M}\over f^-_M c^2_M}D=-{\rig{f_{R^{(+1}}}{T_M}\over c^2_M f^-_M}D+{1\over c_{R^{(+1}}c_M}D {\rig{f_{R^{(+1}}}{T_M}\over f^-_M}-{1\over c_{R^{(+1}}c_M }{\rig{f_{R^{(+1}}}{T_M}\over f^-_M}D
$$ 
It is easy to see that coefficient before terms linear in $D$ equal to zero. 

By the same way it is possible to check that the last expression is equality.  

We present below only calculations the most complicate terms of the third degree in $f^{(-1}$ functions to obtain the value $(R^{(+1},S^{(-1})$ unknown up to now. These terms are the following ones (we conserve in mind common factor ${1\over c_{R^{(+1}}c_M(f^-_M)^3}$, which will be taken into account in the end of computation)
$$
{1\over 2c_M}([X_{-R^{(+1}},X^+_M][c,f^{(-1}])\theta-{1\over 6}([X_{-R^{(+1}},X^+_M][c,[[\alpha^{(+1},f^{(-1}]f^{(-1}]])
$$
These terms arise after separation the third degree terms from $[c,\rig{f^{(-1}}{T_M}]$ (see section Frechet derivative).
$$
-{1\over 2}(X_{-R^{(+1}}[\alpha^{(+1}[c[\alpha^{(+1},f^{(-1}]]-{1\over 2}(X_{-R^{(+1}}[([\alpha^{(+1},f^{(-1}]_h[c,\alpha^{(+1}]]
$$
Two last terms arise from $[c,\rig{f^0}{T_M}]$ and $(X_{-R^{(+1}}[([\alpha^{(+1},f^{(-1}])_h][c,\rig{f^{+1}}{T_M}]])$ correspondingly (about the symbol  $A_h$ see section discrete transformation, below this is very important).
And at last from $(\phi'(f))_{-M+S^{(-1},-M}$ we have in addition
$$
{1\over 2}([X_{-R^{(+1}},X^+_M]f^{(-1})\theta-{c_M\over 6}[[\alpha^{(+1},f^{(-1}]f^{(-1}]
$$
Further transformations are connected with the single relation
$$
[c[[\alpha^{(+1},f^{(-1}]f^{(-1}]]-c_M[[\alpha^{(+1},f^{(-1}]f^{(-1}]=
3[[\alpha^{(+1},f^{(-1}],[c,f^{(-1}]]+3\theta f^{(-1}
$$
which easy follows from the rules of commutation, definition of $\alpha^{(+1}=[X^+_M,f^{(-1}]$ and commutation relation $[f^{(-1}[c,f^{(-1}]]=\theta X^-_M\equiv (\alpha^{(+1}[c,f^{(-1}])X^-_M$.
After some not complicate calculation all terms above lead to finally result 
$$
{c_{R^{(+1}}\over 2c_M}\theta ([X_{-R^{(+1}},X^+_M]f^{(-1})
$$
or after taking into account factors omitted above we come to the finally expression for the 
sum of all terms presented above
$$
{1\over 2(c_M)^2(f^-_M)^2}\theta \rig{f_{R^{(+1}}}{T_M}
$$
Up to now we have not took into account linear in Cartan elements terms in 
$(X_{-R^{(+1}}[([\alpha^{(+1},f^{(-1}])_h][c,\rig{f^{+1}}{T_M}]])$.  
Account of this term lead to result ( Cartan matrix in our consideration is symmetrical)
$$
\sum ([h_{\beta},X_{-R^{(+1}}][c,\rig{f^{+1}}{T_M}])k^{-1}_{\beta,\gamma}(h_{\gamma}[\alpha^{(+1},f^{(-1}])=c_{R^{(+1}}\rig{f_{R^{(+1}}}{T_M}\sum n^{R^{(+1}}_{\gamma} (\alpha^{(+1},[h_{\gamma},f^{(-1}])=
$$
$$
c_{R^{(+1}}\rig{f_{R^{(+1}}}{T_M}\sum (R^{(+1}S^{(-1})f_{S^{(-1}}f_{(S^{(-1})'}(X_{(S^{(-1})'}[X^+_M,X_{S^{(-1}}])
$$
where $(R^{(+1}S^{(-1})=\sum n^{R^{(+1}}_{\gamma}k_{\gamma,\beta}n^{S^{(-1}}_{\gamma}$

Now let us produce some transformation of equivalence with scalar product $\theta_P=([c,f^{(-1}][X^+_M,[\sum (g^i_{P} g^i),f^{(-1}]]))$
$$
\theta_P=\sum_{S^{(-1}} c_{-M-S^{(-1}}(P S^{(-1})f_{-M-S^{(-1}}f_{S^{(-1}}(X_{-M-S^{(-1}}[X^+_M,X_{S^{(-1}}])=
$$
$$
-c_M\sum_{S^{(-1}} (P S^{(-1})f_{-M-S^{(-1}}f_{S^{(-1}}(X_{-M-S^{(-1}}[X^+_M,X_{S^{(-1}}])+
$$
$$
\sum_{S^{(-1}} c_{-M-S^{(-1}}(P -M-S^{(-1})f_{-M-S^{(-1}}f_{S^{(-1}}(X_{-M-S^{(-1}}[X^+_M,X_{S^{(-1}}])=
$$
$$
-c_M\sum_{S^{(-1}} (P S^{(-1})f_{-M-S^{(-1}}f_{S^{(-1}}(X_{-M-S^{(-1}}[X^+_M,X_{S^{(-1}}])-
(PM)\theta - \theta_P.
$$
The last equality allow present $Q$ in a form
$$
\theta_P=-{1\over 2}c_M\sum_{S^{(-1}} (P S^{(-1})f_{-M-S^{(-1}}f_{S^{(-1}}(X_{-M-S^{(-1}}[X^+_M,X_{S^{(-1}}])-{1\over 2}(PM)\theta 
$$
In the case under consideration $P=R^{(+1}, (R^{(+1}M)=1$ and two terms above exactly opposite
the sign with the same calculated before cancels with them and thus main equation is satisfied.
  
\subsubsection{Matrix element between states $<R^{(+1}|,|Q^{(+1}>$}

$$
\sum_{S^{(-1}}\phi'_{R^{(+1},S^{(-1}}J_{S^{(-1},Q^{(+1}}+\phi'_{R^{(+1},-M}J_{-M,Q^{(+1}}=
$$
$$
\rig{f_{R^{(+1}}}{T_M}(-{(MR^{(+1})\over c_M}+{(Q^{(+1}R^{(+1})\over c_{Q^{(+1}}})\rig{f_{Q^{(+1}}}{T_M}
$$
$$
\sum_{S^{(-1}}\rig{J_1}{T_M}_{R^{(+1},S^{(-1}}{c_{S^{(-1}}\over c_{Q^{(+1}}}\phi'(f)_{-S^{(-1},
-Q^{(+1}}+\rig{J_1}{T_M}_{R^{(+1},r_0}{c_{r_0}\over c_{Q^{(+1}}}\phi'_{-r_0,-Q^{(+1}}+
$$
$$
\sum_{P^{(+1}}\rig{J_1}{T_M}_{R^{(+1},P^{(+1}}{c_{P^{(+1}}\over c_{Q^{(+1}}}\phi'(f)_{-P^{(+1},-Q^{(+1}}
$$

Terms quadratical in $f^{(-1}$ arises in left side only from the second term in a form
$$
(-{1\over c_M}+{1\over c_{Q^{(+1}}})\rig{f_{R^{(+1}}}{T_M}\rig{f_{Q^{(+1}}}{T_M}
$$ 
The same structure have two first terms in right side. Summation on $S^{(-1}$ in the first sum
lead to
$$
{c_{P^{(+1}}+c_{Q^{(+1}}-c_M\over 2c_{Q^{(+1}}c_{P^{(+1}}(f^-_M)^2}([X_{-R^{(+1}}[[X^+_M,X_{-Q^{(+1}}]][\alpha^{(+1},f^{(-1}])
$$
Summation on $r_0$ in the second sum lead to (we take not into account linear in Cartan elements terms, it will be done separately later)
$$
{1\over c_{Q^{(+1}}c_{P^{(+1}}(f^-_M)^2}([X_{-R^{(+1}}[[\alpha^{(+1},X_{-Q^{(+1}}]][c,\alpha^{(+1}])\equiv
$$
$$
{c_{P^{(+1}}+c_{Q^{(+1}}\over 2c_{Q^{(+1}}c_{P^{(+1}}(f^-_M)^2}([\alpha^{(+1},X_{-R
^{(+1}}][\alpha^{(+1},X_{-Q^{(+1}}])-
$$
$$
{1\over 2c_{Q^{(+1}}c_{P^{(+1}}(f^-_M)^2}(\alpha^{(+1}[c,f^{(-1}])(X^+_M[X_{-Q^{(+1}},X_{-R^{(+1}}])
$$
Summation on $P^{(+1}$ in the last sum lead to three terms which are containing among quadratical
terms in $\phi'(f)_{-P^{(+1},-Q^{(+1}}$ (see section Frechet derivative). One term is coming with expression with $\delta$ after summation 
$$
{1\over 2c_{Q^{(+1}}c_{P^{(+1}}(f^-_M)^2}(\alpha^{(+1}[c,f^{(-1}])(X^+_M[X_{-Q^{(+1}},X_{-R^{(+1}}])
$$ 
which is exactly opposite by the sign of the last term in line above. Second equal to the next term in this expression and after summation on $P^{(+1}$ takes the form
$$
-{1\over c_{P^{(+1}}}\rig{f_{R^{(+1}}}{T_M}\rig{f_{Q^{(+1}}}{T_M}
$$ 
and the last more complicate term after summation having the form
$$
{c_M\over 2c_{P^{(+1}}c_{Q^{(+1}}(f^-_M)^2}([X_{-R^{(+1}}[[X^+_M,X_{-Q^{(+1}}]][\alpha^{(+1},f^{(-1}])
$$
which cancels last term of the first sum. Now by simple computation we come to equality
$$
([X_{-R^{(+1}}[[X^+_M,X_{-Q^{(+1}}]][\alpha^{(+1},f^{(-1}])=2(f^-_M)^2\rig{f_{R^{(+1}}}{T_M}\rig{f_{Q^{(+1}}}{T_M}-
$$
$$
([\alpha^{(+1},X_{-R^{(+1}}][\alpha^{(+1},X_{-Q^{(+1}}])
$$
All calculated above terms are mutually cancels and now it is necessary to calculate omitted above terms linear in Cartan elements under summation on $r_0$. We have
$$
{1\over c_{P^{(+1}}c_{Q^{(+1}}f^-_M}\sum ([h_{\beta},X_{-Q^{(+1}}][c,\rig{f^{+1}}{T_M}])k^{-1}_{\beta,\gamma}(h_{\gamma}[\alpha^{(+1},X_{-Q^{(+1}}])=
$$
$$
{1\over c_{Q^{(+1}}}\rig{f_{R^{(+1}}}{T_M}\rig{f_{Q^{(+1}}}{T_M}\sum n^{R^{(+1}}_{\gamma} k_{\gamma,\beta}n^{Q^{(+1}}_{\beta}
$$
and finally we come to equalities 
$$
(MR^{(+1})=1,(Q^{(+1}R^{(+1})=-\sum n^{R^{(+1}}_{\gamma} k_{\gamma,\beta}n^{Q^{(+1}}_{\beta}
$$
 which coincide with the proposition of the main assertion.

\subsubsection{Matrix element between states $<r_0|,|M>$}

In this case the main equation looks as
$$
\phi'_{r_0,S^{(-1}}J_{S^{(-1},M}+\phi'_{r_0,-M}J_{-M,M}=
$$
$$
\rig{f^0}{T_M}(-{(r_0,M)\over f^-_M c^2_M}D{1\over f^-_M}+{1\over (f^-_M)^3 c^2_M}\theta_{r_0})+\sum \rig{J^0_1}{T_M}_{r_0,R^{(+1}})((\phi'(f)^{-1})^T)_{R^{(+1},M}+
$$
$$
\sum \rig{J^p_1}{T_M}_{r_0,r'_0}((\phi'(f)^{-1})^T)_{r'_0,M}+\sum \rig{J^0_1}{T_M}_{r_0,S^{(-1}})((\phi'(f)^{-1})^T)_{S^{(-1},M}
$$
We rewrite equation above in more detail form, fulfilling where possible summation 
$$
\sum ({1\over c_{S^{(-1}} c_M}){([\alpha^{(+1},X_{S^{-1}}]X_{-r_0})\over f^-_M}([X_{-S^{-1}},X^-_M][c,f^{(+1}])-
$$
$$
{([\alpha^{(+1},f^{(-1}]X_{-r_0})\over 2(f^-_M)^2 c^2_M}D=\rig{f^0}{T_M}(-{(r_0,M)\over f^-_M c^2_M}D{1\over f^-_M}+{1\over (f^-_M)^3 c^2_M}\theta_{r_0})+
$$
$$
{1\over c_M c_{r_0}}\sum ([X_{-r_0},X_{-R^{+1}}][c,\rig{f^{(+1}}{T_M}])\phi'(f)_{-R^{(+1},-M}-
$$
$$
{1\over 2c_M c_{r_0}}([X_{-r_0}[\alpha^{(+1},f^{(-1}]_h][c,\rig{f^0}{T_M}])+{1\over 2c_M c_{r_0}}D{([\alpha^{(+1},f^{(-1}]X_{-r_0})\over 2(f^-_M)^2}-
$$
$$
{1\over (f^-_M)^2 c_M c_{r_0}}([X_{-r_0},\alpha^{(+1}][c,\rig{f^{-1}}{T_M}]]
$$
The terms linear in $D$ except of written explicit above are presented also in 
$\phi'(f)_{-R^{(+1},-M}$ in a form ${f_{-R^{(+1}}\over f^-_M c_M}D$. Substituting this expression in corresponding place we rewrite it (after not cumbersome manipulations) in the form  
$$ 
{1\over (f^-_M)^2c^2_M c_{r_0}} ([X_{-r_0},f^{(-1}][c,[X^+_M,f^{(-1}]])=
$$
$$
-{c_M+c_{r_0}\over 2(f^-_M)^2c^2_M c_{r_0}}([[X^+_M,f^{(-1}]f^{(-1}]X_{-r_0})
$$

Thus linear in $D$ terms are the following ones
$$
-{1\over 2}{([[X^+_M,f^{(-1}]f^{(-1}]X_{-r_0})\over (f^-_M)^2c^2_M}D...= 
{1\over c_{-r_0}c^2_M f^-_M}([X_{-r_0},f^{(-1}][c,\rig{f^{(+1}}{T_M}])
$$
$$
+{1\over 2c_{-r_0}c_M}D{([[X^+_M,f^{(-1}]f^{(-1}]X_{-r_0})\over (f^-_M)^2}+...
$$
having as its result $(r_0,M)=0$.

Now we would like to compare remaining terms zero degree in $D$ in both sides and find value for $(r_0,S^{-1})$.

First of all necessary compare all nonlocal terms in both sides. Such nonlocal terms arises as result of summation on $S^{(-1}$ in first line and summation on $R^{(+1}$ in the second line.
It is necessary take in mind that only one term in $\phi'(f)_{-R^{(+1},-M}$ is not local namely
${1\over c_{-R^{(+1}}}([X^-_M,X_{R^{(+1}}][c,f^{(+1}])$. Changing in summation $-R^{(+1}\to S^{(-1}$ we obtain
$$
{1\over c_M}\sum ({1\over c_{S^{(-1}}}+{1\over c_{r_0}}){([\alpha^{(+1},X_{S^{-1}}]X_{-r_0})\over f^-_M}([X_{-S^{-1}},X^-_M][c,f^{(+1}])
$$
Nonlocal terms in this sum and in the sum of left side are cancels and result of summation exactly opposite by the sign with the same term arising from $[c,\rig{f^{-1}}{T_M}]$.

By the same technique it is possible to convince that all terms in equation under consideration are mutually cancel. We demonstrate these calculations on two most complicate examples.

At first we calculate  terms containing $\rig{f^0}{T_M}$. Matrix coefficient before before 
$[c,\rig{f^0}{T_M}]$ is the following one ( up to common numerical factors)
$$
-{1\over 2}[X_{-r_0}[\alpha^{(+1},f^{(-1}]_h]+[[X_{-r_0},\alpha^{(+1}]f^{(-1}]=
$$
$$
{1\over 2}([[X_{-r_0},\alpha^{(+1}]f^{(-1}]+[[X_{-r_0},f^{(-1}]\alpha^{(+1}])-
$$
$$
{1\over 2}\sum [X_{-r_0}h_{\beta}]k^{-1}_{\beta,\gamma}(h_{\gamma}[\alpha^{(+1},f^{(-1}]])
$$
Expression in the second line equal $0$. Indeed    
$$
[[X_{-r_0},\alpha^{(+1}]f^{(-1}][c,\rig{f^0}{T_M}]=[[X^+_M [[X_{-r_0},f^{(-1}]f^{(-1}]][c,\rig{f^0}{T_M}]=[\alpha^{(+1}[X_{-r_0},f^{(-1}]][c,\rig{f^0}{T_M}]
$$
After scalar multiplication on $[c,\rig{f^{-1}}{T_M}]$ it is necessary calculate two terms
$$
[X_{-r_0}h_{\beta}][c,\rig{f^0}{T_M}]=c_{r_0}\sum n^{r_0}_{\gamma}k_{\gamma,\beta}\rig{f_{r_0}}{T_M}
$$
$$
(h_{\gamma}[\alpha^{(+1},f^{(-1}]])=\sum_{S^{-1},\beta} n^{S^{-1}}_{\beta}k_{\gamma,\beta}
f_{S^{-1}}f_{-M-S^{-1}}([X^+_M,X_{-M-S^{-1}}]X_{S^{-1}})
$$
Substituting these expressions we come to finally result
$$
\rig{f_{r_0}}{T_M}{1\over 2c_M (f^-_M)^3}\sum_{S^{-1}} (S^{-1}r_0)f_{S^{-1}}f_{-M-S^{-1}}([X^+_M,X_{-M-S^{-1}}]X_{S^{-1}})=-
$$
$$
{1\over (c_M)^2(f^-_M)^3}\rig{f_{r_0}}{T_M}\theta_{r_0}
$$ 
with $(S^{-1}r_0)=-\sum n^{S^{-1}}_{\beta}k_{\beta,\gamma}n^{r_0}_{\gamma}$.
The last equality written in connection with general formulae in the end of the section $(6.2.4)$ in which $(M r_0)=0$. 

Now we calculate terms of fourth  degree in $f^{(-1}$ functions. They all arises as scalar product with generator $X_{-r_0}$. Up to common factor ${1\over c_{-r_0}c_M(f^-_M)^3}$ they are the following ones 
$$
-{1\over 6}[[[\alpha^{(+1},f^{(-1}]f^{(-1}][c,\alpha^{(+1}]]-{1\over 6}[\alpha^{(+1}[c[[\alpha^{(+1},f^{(-1}]f^{(-1}]]]-
$$
$$
{1\over 2}[\alpha^{(+1}[f^{(-1}[c[\alpha^{(+1},f^{(-1}]]]]
$$
The first two terms arises after summation in corresponding terms of $\phi'(f)_{-R^{(+1},-M}$ and 
$[c,\rig{f^{-1}}{T_M}]$. The last one after representation $f^0=\rig{f^0}{T_M}-{1\over 2}[\alpha^{(+1},f^{(-1}]$ in $[c,\rig{f^{-1}}{T_M}]$. Except of these 3 terms fourth degree $f^{(-1}$ have two terms containing $\theta$ function. We present 3 terms above to the same form after which cancels all terms of fourth order became obvious.
For this goal let us consider obvious equality
$$ 
[[[\alpha^{(+1},f^{(-1}]f^{(-1}]f^{(-1}]=-([\alpha^{(+1},f^{(-1}][\alpha^{(+1},f^{(-1}])X^-_M
$$
and perform commutation both sides with $X^+_M$. We obtain
$$
[[[\alpha^{(+1},f^{(-1}]\alpha^{(+1}]f^{(-1}]+[[[\alpha^{(+1},f^{(-1}]f^{(-1}]\alpha^{(+1}]=
2[[[\alpha^{(+1},f^{(-1}]\alpha^{(+1}]f^{(-1}]=
$$
$$
-([\alpha^{(+1},f^{(-1}][\alpha^{(+1},f^{(-1}])H
$$
From the last result it follows that the both terms above with coefficient ${1\over 6}$ are equal to each other $((X_{-r_0} H)=0)$ and after application to this term equality for
$[c[[\alpha^{(+1},f^{(-1}]f^{(-1}]]]$ ( see middle of the page 15) we rewrite this term in a
form 
$$
-[\alpha^{(+1}[[\alpha^{(+1},f^{(-1}][c,f^{(-1}]]-[\alpha^{(+1},f^{(-1}]\theta-
{1\over 2}[\alpha^{(+1}[f^{(-1}[c[\alpha^{(+1},f^{(-1}]]]]=
$$
$$
{1\over 2}[\alpha^{(+1},f^{(-1}]\theta
$$
The same structure have two terms arises after summation in corresponding terms of 
$\phi'(f)_{-R^{(+1},-M}$ and in $[c,\rig{f^{-1}}{T_M}]$. Sum of these terms (with the same common factor as above) are the following
$$
{1\over 2c_M}([\alpha^{(+1}[c,f^{(-1}]]+[f^{(-1}[c,\alpha^{(+1}]])\theta=-
{1\over 2}([\alpha^{(+1},f^{(-1}]\theta
$$ 
exactly opposite by the sign with the result obtained few lines above.

This result prove validity matrix element under consideration of the main equation. 

\subsection{Checking of the elements of $(1+N^1+N^0\times 1+N^1)$ rectangular matrix of right dawn corner of (\ref{JHF})}

These calculations are absolutely of the same kind as performed in the previous section. We would like to demonstrate them on detail calculations on example of matrix element 
$<-M|,|S^{(-1}>$.
$$
\phi'_{-M,M}J_{M,S^{(-1}}+\sum_{R^{(+1}} \phi'_{-M,R^{(+1}}J_{R^{(+1},S^{(-1}}+\sum_{r_0}\phi'_{-M,r_0}J_{r_0,S^{(-1}}+\sum_{Q^{(-1}} \phi'_{-M,Q^{(-1}}J_{Q^{(-1},S^{(-1}}
$$
$$
{c_M\over c_{S^{(-1}}}\rig{J_1}{T_M}_{-M,M}\phi'(f)_{-M,-S^{(-1}}-{1\over c_{S^{(-1}}c_M}\sum_{R^{(+1}} \rig{J_1}{T_M}_{-M,R^{(+1}}\phi'_{-R^{(+1},-S^{(-1}}
$$

\subsubsection{Matrix element between states $<-M|,|M>$}

In this case both additional terms arising from $J^{-1}_1$ (see subsection 5.2) must be taken into account. Equation to be checked is as follows
$$
\sum_{S^{(-1}} \phi'_{-M,S^{(-1}}J_{S^{(-1},M}+\phi'_{-M,-M}J_{-M,M}+{(M,M) f^-_M\over c_M^2}D f^+_M - {f^+_M\over c_M^2}\Theta_M=
$$
$$
\sum_{R^{(+1}} \rig{J_1}{T_M}_{-M,R^{(+1}}\phi'_{-R^{(+1},-M}
+\rig{J_1}{T_M}_{-M,M}\phi'(f)_{-M,-M}+{(M,M)\rig{ f^-_M}{T_M}\over c_M^2}D
{1\over f^-_M}+ {\rig{ f^-_M}{T_M}\over c_M^2}\Theta_{-M}
$$
This expression  contain terms of the zero, first and  third  degree of $D$. 
We will present below  only checking terms of the positive degree. $J_{-M,M}=J_{M,-M}={1\over c_M^2}D$ two terms in both sides of the last expression containing this operation leads to (all terms linear in $D$ we present in the left side of equation)
$-[D,\phi'_{-M,-M}]$ and terms in $\phi'_{-M,-M}$ with operation of $D$ are the following ones (see section $4.1$)m${1\over c_M^2}D^2-2{(f^-_M)'-{1\over 2}\Theta\over f^-_Mc_M^2}D$ and thus commutator with $D$ leads to linear term  $-[D,\phi'_{-M,-M}]=2({(f^-_M)'-{1\over 2}\Theta\over f^-_Mc_M^2})'D$.
Linear in $D$ terms in $\phi'_{-M,S^{(-1}}$ after summation lead to ${(f^{-1}[c,f^{+1}])\over c_M^3})$, the same with respect to term $\phi'_{-R^{(+1},-M}$ looks as $ -{([X^+_M ,f^{-1}][c,\rig{ f^{-1}}{T_M}])\over f^-_M c_M^3}$ and thus coefficient before $D$ looks as
$$ 
2({(f^-_M)'-{1\over 2}\Theta\over f^-_M c_M^4})'+{2\over c_M^4}(f^-_M f^+_M)-{(f^{-1}[c,f^{+1}])\over c_M^3})+{([X^+_M ,f^{-1}][c,\rig{ f^{-1}}{T_M}])\over f^-_M c_M^3}-
$$
$$
{2\over c_M^2 f^-_M}{\rig{f^-_M}{T_M}\over f^-_M}
$$
Using equations of discrete transformation and explicit expressions for Frechet derivatives
it is convice that the last expression equal to zero.

\subsubsection{Matrix element between states $<S^{(-1}|,|R^{(+1}>$}

This matrix element is the most difficult for calculations. We will present general formulae below and give some advice in what order this not trivial calculations must be performed.
Terms in basic equation (\ref{JHF}) (in both sides) are the following ones
$$
\sum_{Q^{(+1}} \phi'_{S^{(-1},Q^{(+1}}J_{Q^{(+1},R^{(+1}}+\sum_{r_0} \phi'_{S^{(-1},r_0}J_{r_0,R^{(+1}}+\sum_{Q^{(-1}} \phi'_{S^{(-1},Q^{(-1}}J_{Q^{(-1},R^{(+1}}+
$$
$$
\phi'_{S^{(-1},-M}J_{-M,R^{(+1}}+f_{R^{(+1}}({(R^{(+1},M) \over c_M} - {(R^{(+1},S^{(-1})\over c_{S^{(-1}}})f_{S^{(-1}}=
$$
$$
\rig{f_{S^{(-1}}}{T_M}({(R^{(+1},M) \over c_M} - {(R^{(+1},S^{(-1})\over c_{R^{(+1}}})
\rig{f_{R^{(+1}}}{T_M}+\rig{J}{T_M}_{S^{(-1},M}{c_M\over c_{R^{(+1}}}\phi'_{-M,-R^{(+1}}+
$$
$$
\sum_{Q^{(+1}} \rig{J}{T_M}_{S^{(-1},Q^{(+1}}{c_{Q^{(+1}}\over c_{R^{(+1}}}
\phi'_{-Q^{(+1},-R^{(+1}}+\sum_{r_0} \rig{J}{T_M}_{S^{(-1},r_0}{c_{r_0}\over c_{R^{(+1}}}
\phi'_{-r_0,-R^{(+1}}+
$$
$$
\sum_{Q^{(-1}} \rig{J}{T_M}_{S^{(-1},Q^{(-1}}{c_{Q^{(-1}}\over c_{R^{(+1}}}
\phi'_{-Q^{(-1},-R^{(+1}}
$$
Now we present the same formulae after summation in which it is only necessary substitute explicit expressions for Frechet derivatives and transformed by discrete transformation values
$\rig{f}{T_M}_P$ (all terms are written in the same order as above)
$$
{f^-_M f^+_M\over c_{S^{(-1}} c_{R^{(+1}}}\delta_{S^{(-1}, -R^{(+1}}-{1\over c_{S^{(-1}} c_{R^{(+1}}}([[X_{-S^{(-1}},f^{(-1}]_h X_{-R^{(+1}}][c,f^{(+1}])+
$$
$$
{1\over c^2_{R^{(+1}}}\phi'_{S^{(-1},-R^{(+1}}D+{1\over c_{R^{(+1}}}\sum_{Q^{(-1}} \phi'_{S^{(-1},Q^{(-1}}{1\over c_{Q^{(-1}}}([X_{-Q^{(-1}},X_{-R^{(+1}}][c,f^0])+
$$
$$
-{1\over c_{R^{(+1}}c_M}\phi'_{S^{(-1},-M}([X^+_M,X_{-R^{(+1}}][c,f^{(-1}])+f_{R^{(+1}}({(R^{(+1}M) \over c_M} - {(R^{(+1}S^{(-1})\over c_{S^{(-1}}})f_{S^{(-1}}=
$$
$$
\rig{f_{S^{(-1}}}{T_M}({(R^{(+1}M) \over c_M} - {(R^{(+1}S^{(-1})\over c_{R^{(+1}}})
\rig{f_{R^{(+1}}}{T_M}+{1\over c_{S^{(-1}} c_{R^{(+1}}}([[X_{-S^{(-1}},X^-_M][c,\rig{f^{(+1}}{T_M}])\phi'_{-M,-R^{(+1}}-
$$
$$
{1\over c_{S^{(-1}} c_{R^{(+1}}}D\phi'_{S^{(-1},-R^{(+1}}+{1\over c_{S^{(-1}} c_{R^{(+1}}}
\sum_{Q^{(+1}} ([[X_{-S^{(-1}},X_{-Q^{(+1}}][c,\rig{f^0}{T_M}])\phi'_{-Q^{(+1},-R^{(+1}}
$$
$$
{1\over c_{S^{(-1}} c_{R^{(+1}}}([[X_{-S^{(-1}}[\alpha^{(+1},X_{-R^{(+1}}]_h][c,\rig{f^{(-1}}{T_M}])+{c_M\rig{f^-_M}{T_M}\over c_{S^{(-1}} c_{R^{(+1}}f^-_M}\delta_{S^{(-1}, -R^{(+1}}
$$
Now we would like to show that all non locality in terms above are mutually cancel. 
Let us rewrite two last terms of $\phi'_{S^{(-1},Q^{(-1}}$ in equivalent form
$$
-{1\over c_{S^{(-1}}}(X_{-S^{(-1}}[X_{Q^{(-1}}[c,f^0]])+
{1\over 2f^-_M}(X_{-S^{(-1}}[[\alpha^{(+1},f^{(-1}]X_{(Q^{(-1}}])=
$$
$$
-{1\over c_{S^{(-1}}}(X_{-S^{(-1}}[X_{Q^{(-1}}[c,\rig{f^0}{T_M}]]])+
{c_{(Q^{(-1}}\over 2f^-_M}(X_{-S^{(-1}}[[\alpha^{(+1},f^{(-1}]X_{(Q^{(-1}}])
$$
From which follows only one non local term in the sum in the left side has the form
$$
-{1\over c_{S^{(-1}} c_{R^{(+1}}}\sum_{Q^{(-1}} {1\over c_{Q^{(-1}}}([X_{-Q^{(-1}},X_{-R^{(+1}}][c,f^0])(X_{-S^{(-1}}[X_{Q^{(-1}}[c,\rig{f^0}{T_M}]]])
$$
Exact the same is non local term in the right hand side of sum. All other terms allow summation and after substitution. After in principle non cumbersome calculations -- it is necessary compare coefficients before terms of the same structure -- we convince that expression written above is equality. 
  
We will not present checking all other matrix elements because the tremendous number of cancelation under calculations above can not be occasional.   

\section{Outlook}

The main positive result of the present paper - two Poisson structures in explicit form invariant with respect to discrete transformation of the maximal root of arbitrary semi simple algebra. The knowledge of these structures allow to construct equations of hierarchies of n-waves systems in the case of arbitrary semi simple algebra. And thus in this problem this paper put the finally point.

But by the author opinion the most interesting is the standing by this paper the following problem
What is group representation background of the discrete transformation of maximal root?
All calculations above were done using only main properties of the structure theory of semi simple algebra (properties  of its  root space and commutation relation in it). Discrete substitution is some new element of this construction possible taken ad hoc from the head. But it properties can be proofed only on the base of group representation theory and thus it must
have some group theoretical origin. What? This is unsolved in the present paper very interesting problem.    

Keeping in mind that discrete transformation of maximal root is consequence of Yang-Mills equations describing interaction of elementary particles after finding the answer on the stated above problem it will be possible to understand why equations of Y-M are excluded from other ones and why Great God have used them in the process of construction of our Universe.

Some comments about perspectives of possible further investigation. It is absolutely clear that it is possible to resolve symmetry equation in the case of four dimensional self dual system and by this way construct all integrable systems in the sense of Richard Ward. The same is true for all super symmetrical generalization of Y-M self dual equations.
  
Author hope return to solution of this problem in the nearest time.

\section{Acknowledgments}

Author thanks Jorge Uruchurtu Chavarin by the pen of whom this paper was began for many stimulating, non formal discussions and CONNECUT for finance support.

\section{Appendix}

The terms with derivatives in (\ref{FUNK}) are the following ones
\begin{equation}
{1\over 2c_M}({X'_M\over X_M})^2+{1\over c_M f^-_M}([X^+_M,f^{(-1}][c,(f^{(-1})'])\label{FUHK1}
\end{equation}
and thus it is necessary to check connection of following mixed derivatives
$$
G^T_{M,M}-G_{M,M}, \quad G^T_{M,f^{(-1}}-G_{f^{(-1},M},\quad G^T_{f^{(-1},f^{(-1}}-G_{f^{(-1},f^{(-1}}
$$
Consider the first difference under arbitrary dependence of generating functional on $(x,x')$
variables.
$$
\frac{\delta G}{\delta x}=-G'_{x'}+G_x=-x''G'_{x',x'}-x'G'_{x',x}+G_x,
$$
$$
\frac{\delta^2 G}{\delta x,\delta x}=-G'_{x',x'}D^2+(-x''G'_{x',x'}-x'G'_{x',x}+G_x)_{x'}D+
(-x''G'_{x',x'}-x'G'_{x',x}+G_x)_x
$$ 
Thus
$$
G^T_{x,x}-G_{x,x}=-D^2G_{x',x'}+G_{x',x'}D^2-2D(-x''G_{x',x'}-x'G_{x',x}+G_x)_{x'}+(x''G_{x',x',x'}+x'G_{x',x',x})'=
$$
$$
-2G'_{x',x'}D-(G_{x',x'})''-2(-x''G_{x',x',x'}-x'G_{x',x',x})D-(-x''G_{x',x',x'}-x'G_{x',x',x})'=0
$$
Calculations and comparing derivatives of the second order connected with last term in (\ref{FUHK1})
$$
G_{f^{(-1}_i}=-({1\over f^-_M}([X^+_M,f^{(-1}][c,X^-_i])'+{1\over f^-_M}([X^+_M,X^-_i][c,(f^{(-1})'])=
$$
$$
-({1\over f^-_M})'([X^+_M,f^{(-1}][c,X^-_i]-{c_M\over f^-_M}([X^+_M,X^-_i](f^{(-1})')
$$
$$
G_{f^{(-1}_i,f^{(-1}_j}=-({1\over f^-_M})'([X^+_M,X^-_j][c,X^-_i]-{c_M\over f^-_M}(X^+_M[X^-_i,X^-_j])D
$$
From which equality $G_{f^{(-1}_i,f^{(-1}_j}=G_{f^{(-1}_j,f^{(-1}_i}^T$ ($(AD)^T=-AD$ ! follows immediately.
The same calculations lead to equality $G^T_{M,f^{(-1}}=G_{f^{(-1},M}$.

\end{document}